\documentclass[12pt]{JHEP3}

\usepackage{amssymb}
\usepackage{amsbsy}
\usepackage{graphicx}

\newcommand{\be}{\begin{equation}} \newcommand{\ee}{\end{equation}}
\newcommand{\bea}{\begin{eqnarray}} \newcommand{\eea}{\end{eqnarray}}
\newcommand{\el}{\nonumber \\}
\newcommand{\re}[1]{(\ref{#1})}

\newcommand{\adot}{\dot{a}}
\newcommand{\addot}{\ddot{a}}
\newcommand{\bdot}{\dot{b}}
\newcommand{\bddot}{\ddot{b}}

\newcommand{\rhodot}{\dot{\rho}}
\newcommand{\Hdot}{\dot{H}}

\newcommand{\brt}[1]{[#1]}
\newcommand{\rmd}{\textrm{d}}
\newcommand{\ack}{\acknowledgments}
\renewcommand{\a}{\alpha}
\renewcommand{\b}{\beta}
\renewcommand{\c}{\gamma}
\renewcommand{\d}{\delta}

\renewcommand{\H}{\frac{\adot}{a}}
\newcommand{\HH}{\frac{\adot^2}{a^2}}
\newcommand{\HHH}{\frac{\adot^3}{a^3}}

\newcommand{\Hb}{\frac{\bdot}{b}}
\newcommand{\HHb}{\frac{\bdot^2}{b^2}}
\newcommand{\HHHb}{\frac{\bdot^3}{b^3}}
\newcommand{\HHHHb}{\frac{\bdot^4}{b^4}}
\newcommand{\Hd}{\frac{\addot}{a}}
\newcommand{\Hbd}{\frac{\bddot}{b}}

\newcommand{\PRD}[1]{{\it Phys. Rev.} {\bf D#1}}
\newcommand{\PRL}[1]{{\it Phys. Rev. Lett.} {\bf #1}}

\newcommand{\NPB}[1]{{\it Nucl. Phys.} {\bf B#1}}
\newcommand{\PLB}[1]{{\it Phys. Lett.} {\bf B#1}}

\newcommand{\APJ}[1]{{\it Astrophys. J.} {\bf #1}}

\newcommand{\JHEP}[1]{{\it JHEP} {\bf #1}}
\newcommand{\CQG}[1]{{\it Class. Quant. Grav.} {\bf #1}}
\newcommand{\GRG}[1]{{\it Gen. Rel. Grav.} {\bf #1}}

\title{Lovelock inflation and the number of large dimensions}

\author{Francesc Ferrer \\
CERCA, Department of Physics, Case Western Reserve University,
10900 Euclid Avenue, Cleveland, OH 44106-7079, USA \\
Email: \email{francesc {\it dot} ferrer {\it at} case {\it dot} edu}}

\author{Syksy R\"{a}s\"{a}nen \\
CERN, Physics Department, Theory Unit,
CH-1211 Geneva 23, Switzerland \\
Email: \email{syksy {\it dot} rasanen {\it at} iki {\it dot} fi}}

\abstract{We discuss an inflationary scenario based on Lovelock terms.
These higher order curvature terms can lead to inflation
when there are more than three spatial dimensions.
Inflation will end if the extra dimensions are stabilised, so that
at most three dimensions are free to expand. This relates graceful
exit to the number of large dimensions.}


\preprint{CERN-PH-TH/2007-113}

\begin{document}

\setcounter{secnumdepth}{3}

\section{Introduction}

\paragraph{Compactification and inflation.}

Perturbative string theory is most naturally formulated in
9+1 dimensions. The usual way of getting closer to the observed
3+1-dimensional universe is to compactify six spatial dimensions
by hand so as to end up with four-dimensional Minkowski space
times a six-dimensional compact manifold. (It is also possible to
formulate critical string theory without extra spacetime dimensions;
see e.g. the review \cite{Horowitz:2004}.)

The extra dimensions are usually taken to be static (indeed,
understanding of string theory in time-dependent backgrounds is
still quite limited), and compactification is
considered not to involve any dynamical evolution.
In the search for a static split into large and small
spatial dimensions, no explanation has emerged for why
there should be three of the former and six of the latter. From the point
of view of the ten-dimensional theory, there is no particular
preference for six compact dimensions.

Even if the 3+6 split is taken for granted, there is a vast number
of different ways of compactifying the six dimensions.
Thus far, no unique, or even uniquely promising, compactification
has emerged, and it has been suggested that there
simply is no preferred way to compactify the extra dimensions.
This could indicate a lack of predictivity in string theory
(or that string theory is not correct), but it may rather show
that some important principle is missing. There is no complete
non-perturbative formulation of string theory, and it could be
that the required ingredient is related to poorly understood
non-perturbative aspects. A simpler possibility is that
the split into three large and six small dimensions
arises due to dynamical evolution, which is absent in
the usual formulations of the problem, based as they are on a
particle physics viewpoint with static manifolds, rather than a
cosmological approach with evolving dimensions.

A somewhat analogous situation existed with respect to the puzzle
of cosmological homogeneity and isotropy before the introduction of
inflation. General relativity has a multitude of solutions,
and though no rigorous measure in the space of solutions has been
found, it would seem that the homogeneous and isotropic
Friedmann-Robertson-Walker (FRW) models are a subset of
measure zero by any reasonable definition.
So the question is: why is the universe, for a large segment of its
evolution, well described by one of these very special solutions?
Attempts to solve this problem in the context of general relativity
\cite{Misner} were unsuccessful until the introduction of
inflation using ideas from particle physics \cite{Guth:1981}.
From a modern viewpoint, the solution of the problem by
accelerating expansion is intimately related to violation of the
dominant energy condition ($\rho+3p\geq0$), an ingredient which may
seem strange from a general relativistic point of view, but which is natural
in particle physics. (However, it is not clear how generally inflation
can start and lead to homogeneity and isotropy from an
inhomogeneous and anisotropic initial state \cite{Trodden, Ellis}.)

Perhaps taking into account the ingredient of dynamical
evolution, which is natural from the cosmological point of view,
could similarly be useful with the particle physics problem of
compactification. 
At late times in the universe, the visible spatial dimensions expand,
while any compact dimensions must be
relatively static, so as not to conflict with the observational
limits on the change of the gravitational coupling
(see for example \cite{Ferrer:2005}).
From the cosmological point of view, the question is: which
mechanism is responsible for stabilising some of the dimensions
while others are free to expand, and how does that mechanism
determine the number of expanding dimensions?

Though compactification is a well-studied topic, relatively
little work has been done on trying to explain why the number of
large spatial dimensions should be three.
The most notable exception is the study of
string gas cosmology (SGC), where the dynamical determination of
the number of large dimensions has been a central topic
\cite{Kripfganz:1988, Brandenberger:1989, Tseytlin:1991}
(see \cite{Battefeld:2005, Brandenberger:2007} for reviews).
(There is also an alternative explanation for why we observe
three large dimensions: that we live on a three-dimensional brane.
There has been some work on trying to dynamically determine
why three-branes would be preferred in this case \cite{Durrer, Karch:2005}.)

In SGC, all spatial dimensions start on an
equal footing, all compact and of the string size.
The universe is filled with a hot gas of branes of
all allowed dimensionalities. In the simplest versions of SGC the
dimensions are toroidal, so that branes can wind around them, and
resist expansion. (If the particle physics compactifications
are unmotivated from a cosmological point of view, toroidal extra
dimensions are in turn problematic for particle physics.
See \cite{Easson:2001, Easther:2002a} for
discussion of more complex compactifications.)
As the universe expands and cools down, winding and anti-winding
modes annihilate, allowing further expansion.
A simple counting argument suggests that $p$-branes
and their anti-branes cannot find each other to annihilate in more
than $2p+1$ spatial dimensions, so at most $2p+1$ dimensions
can become large. For $p=1$, corresponding to strings, this is three
spatial dimensions. (Some quantitative studies of brane gases have
cast doubt on this qualitative argument, see
\cite{Sakellariadou:1995, Easther:2002b, Bassett:2003, Easther:2003, Easther:2004, Danos:2004}
for different analyses.)

Conceptually, inflation fits naturally into SGC: all dimensions
are initially small, and inflation makes three of them macroscopically
large. Instead of having separately inflation in the visible dimensions
and static compactification in the extra dimensions, one could
dynamically explain decompactification via inflation.
(This idea was introduced in an earlier Kaluza-Klein context
in \cite{Shafi}.) However,
the practical implementation of inflation in SGC is problematic,
since inflation dilutes the string gas which stabilises the extra
dimensions, and no compelling inflationary scenario in SGC has been found
\cite{Brandenberger:2003, Patil:2004, Kaya:2004, Biswas:2005, Easson:2005, Shuhmaher:2005, Battefeld:2006}.
(For alternatives to inflation in SGC, see
\cite{Nayeri:2005, Brandenberger:2005, Brandenberger:2006, Kaloper:2006}.)
An extra ingredient is needed, something that stabilises the
extra dimensions even against inflation.
We will point out that if such a mechanism is found, stabilising
the extra dimensions may be directly related to ending inflation
in the visible dimensions.

\paragraph{Lovelock gravity.}

We are interested in inflation in a higher-dimensional
space. In a general metric theory of gravity in $d$ dimensions,
the equation of motion sets the energy-momentum tensor equal to some 
covariantly conserved rank two tensor built from the
metric and its derivatives. Demanding the equations of motion to be
of second order \cite{Muller-Hoissen:1991, Simon:1990, Woodard:2006}
strongly constrains the terms which can appear. In four dimensions, there are
only two local tensors with the required properties: the Einstein
tensor, and the metric itself, the latter corresponding to the
cosmological constant \cite{Lovelock:1971, Lovelock:1972}.

In more than four dimensions, the Einstein tensor is no longer the
unique covariantly conserved non-trivial tensor constructed from the
metric and its first and second derivatives. In $d$ dimensions there
are exactly $[d/2]$ ($d/2$ rounded up) such symmetric tensors (and
corresponding local Lagrange densities), known as the Lovelock tensors
\cite{Lovelock:1971}. (The Einstein tensor is still the only covariantly
conserved local tensor which is linear in second derivatives.)

The approach which leads to Einstein gravity in four dimensions
gives Lovelock gravity in higher dimensions.
The first new contribution to the Lagrange density, quadratic in
curvature, is the well-known Gauss-Bonnet term.
In four dimensions it reduces to a topological
quantity and does not contribute to the equations of motion.
(The higher Lovelock terms vanish in four dimensions.)

From the viewpoint of string theory, the Lovelock Lagrangians
may be said to be preferred, as they lead to a unitary and
ghost-free low energy effective theory \cite{Zwiebach:1985, Zumino:1986}.
However, since the effective theory is defined only up to field
redefinitions, Lovelock Lagrangians should be (at least to
second order in the Riemann tensor) physically equivalent to
non-Lovelock Lagrangians \cite{Deser:1986}. This means that
the seeming problems of non-Lovelock terms are expected to
become apparent only at large curvatures, where the effective
theory does not apply.

We do not consider the details of the string theory context,
and will simply look at ten-dimensional cosmology with Lovelock gravity.
From the string theory point of view, we are ignoring the extra
fields present in addition to the metric; in particular we are
assuming that the dilaton has been stabilised in a manner that does
not impose any constraints on the metric.
We find that Lovelock gravity can naturally involve inflation in higher
dimensions. Furthermore, the end of inflation is tied up with the
stabilisation of the hidden dimensions: if the extra dimensions are
kept small, the universe soon becomes effectively four-dimensional.
This will in turn end inflation in the visible dimensions,
because the contribution of the Lovelock terms vanishes in
four dimensions: graceful exit from inflation is tied
to (at most) three spatial dimensions becoming large.

In section 2 we describe Lovelock gravity, explain the inflationary
mechanism and point out the connection between graceful exit and
stabilisation. We briefly discuss some ideas for ending inflation
and summarise in section 3.

\section{Lovelock inflation}

\paragraph{The action and the equation of motion.}

In a metric theory of gravity in $d$ dimensions, the most general
local Lagrange density which leads to equations of motion containing
at most second order derivatives of the metric is \cite{Lovelock:1971}
\bea \label{Lagrangian}
  L_{\mathrm{love}} &=& \sum_{n=0}^{[d/2]} c_n L_n \el
  &\equiv& \sum_{n=0}^{[d/2]} c_n 2^{-n} \delta^{\a_1\cdots\a_{2 n}}_{\b_1\cdots\b_{2 n}} R^{\ \ \ \ \b_1\b_{2}}_{\a_1\a_2} \ldots R^{\ \ \ \ \ \ \ \ \ \b_{2 n - 1}\b_{2 n}}_{\a_{2 n - 1}\a_{2 n}} \ ,
\eea

\noindent where $\delta^{\a_1\cdots\a_k}_{\b_1\cdots\b_k}$
is the generalised Kronecker delta symbol (totally antisymmetric in
both upper and lower indices), $[d/2]$ is $d/2$ rounded up
to the nearest integer and $c_n$ are constants; by definition
$L_0\equiv1$. The first term is the cosmological constant,
the second is the Einstein-Hilbert Lagrange density and the third
is the Gauss-Bonnet Lagrange density. We will consider the case $d=10$,
but for simplicity we drop the terms or order three and four in the
Riemann tensor; including them is straightforward. The action is
\bea \label{action1}
  S_{\mathrm{love}} &=& \int\rmd^{10}x \sqrt{-g} \left( c_0 L_0 + c_1 L_1 + c_2 L_2  \right) + S_\mathrm{m} \el
  &=& \frac{1}{2\kappa^2}\int\rmd^{10}x \sqrt{-g} \left[ - 2 \Lambda + R + \alpha ( R^2 - 4 R_{\a\b} R^{\a\b} + R_{\a\b\c\d} R^{\a\b\c\d} ) \right] \el
  && + \int\rmd^{10}x \sqrt{-g} L_\mathrm{m} \ ,
\eea

\noindent where $L_\mathrm{m}$ is the Lagrangian of the matter fields
present and we have denoted $c_0=-\Lambda/\kappa^2$,
$c_1=1/(2\kappa^2)$ and $c_2=\alpha/(2\kappa^2)$,
where $\Lambda$ is the cosmological constant, $\kappa^2$
is the 10-dimensional gravitational coupling and $\alpha$ is
the Gauss-Bonnet coefficient.

The equation of motion following from \re{action1} is
\bea \label{eom}
  \kappa^2 T_{\mu\nu} = G_{\mu\nu} + \alpha H_{\mu\nu} \ ,
\eea

\noindent where $\kappa^2$ is the gravitational coupling in $d$
dimensions, $T_{\mu\nu}$ is the energy-momentum tensor
(which we take to include the cosmological constant),
$G_{\mu\nu}$ is the Einstein tensor and $H_{\mu\nu}$ is
the Gauss-Bonnet tensor given by
\bea
  H_{\mu\nu} &=& 2 R R_{\mu\nu} - 4 R_{\mu\a} R^{\a}_{\ \nu} - 4 R_{\a\b} R^{\a \ \b}_{\ \mu \ \nu} + 2 R_{\mu\a\b\c} R_{\nu}^{\ \a\b\c} \el
  && - \frac{1}{2} g_{\mu\nu} \left( R^2 - 4 R_{\a\b} R^{\a\b} + R_{\a\b\c\d} R^{\a\b\c\d} \right) \ .
\eea

\paragraph{The metric.}

We take the metric to be the simplest generalisation of the
spatially flat Friedmann-Robertson-Walker (FRW) universe, homogeneous
and separately isotropic in the visible and the extra dimensions:
\bea \label{metric}
  \rmd s^2 = - \rmd t^2 + a(t)^2 \sum_{i=1}^3 \rmd x^{i}\rmd x^{i} + b(t)^2 \sum_{j=1}^6 \rmd y^{j}\rmd y^{j} \ ,
\eea

\noindent where $x^i$ and $y^j$ are the spatial coordinates
in the visible and extra dimensions, respectively.

Given the symmetries of the metric \re{metric}, the energy-momentum
tensor is
\bea \label{emt}
  T^{\mu}_{\ \nu} = \textrm{diag}( -\rho(t), p(t), p(t), p(t), P(t), P(t),  P(t),  P(t),  P(t),  P(t) ) \ .
\eea

With \re{metric} and \re{emt}, the equation of motion \re{eom} reads
\bea
  \label{00} \kappa^2\rho = 3 \HH + 18 \H\Hb + 15 \HHb + 36 \alpha \Hb \left(
2 \HHH + 15 \HH\Hb + 20 \H \HHb + 5 \HHHb \right) \\
  \label{kk} \kappa^2 p = - \left( 2 \Hd + 6 \Hbd + \HH + 12 \H\Hb + 15 \HHb \right) - 12 \alpha\left( 4\H\Hb\Hd + 10 \HHb\Hd \right. \el
   \left. + 2 \HH\Hbd + 20 \H\Hb\Hbd + 20 \HHb\Hbd + 15 \HH\HHb + 40 \H\HHHb + 15 \HHHHb \right) \\
  \label{yy} \kappa^2 ( \rho - 3 p + 2 P ) = 8 \Hbd + 24 \H\Hb + 40 \HHb + 24 \alpha \left( - \HH\Hd - 4 \H\Hb\Hd + 5 \HHb\Hd \right. \el
   \left. - 2 \HH\Hbd + 10 \H\Hb\Hbd + 20 \HHb\Hbd -2 \HHH\Hb + 15 \HH\HHb + 60 \H\HHHb + 25 \HHHHb \right) \ .
\eea

\noindent As in the usual FRW case, not all of the
equations are independent, and (as long as $\bdot\neq0$)
we can simply use \re{00} and \re{kk} along with the conservation
law of the energy-momentum tensor:
\bea \label{cons}
  \rhodot + 3 \H (\rho+p) + 6 \Hb (\rho+P) = 0 \ .
\eea

When the extra dimensions are static, $\bdot=0$, the components of
the Gauss-Bonnet tensor in the four visible directions
vanish, and we recover the usual FRW equations in the visible
directions. This is expected, since in four dimensions the
Gauss-Bonnet term does not contribute to the equations of motion.
Note that the components of the Gauss-Bonnet tensor in the direction
of the extra dimensions do not vanish, though its contribution is
negligible at low curvatures. The higher order Lovelock tensors
vanish when $\bdot=0$ (the expressions for them can be found in
\cite{Demaret:1990}),
so if we used them instead of the Gauss-Bonnet term, the dynamics would
completely reduce to the FRW case when the extra dimensions are stabilised. 
This is presumably related to the fact that in four dimensions the
Gauss-Bonnet action is total derivative, while the higher order Lovelock
actions are identically zero.
For discussion of cosmology with Lovelock terms, see 
\cite{Demaret:1990, Muller-Hoissen:1986, Kripfganz:1987, Deruelle:1990}.

\paragraph{Inflation.}

Let us first look at the case when there is no distinction between the
visible and extra dimensions, so the universe is isotropic, $a=b$.
Then \re{00}--\re{cons} reduce to
\bea
  \label{Hubble} 36 H^2 + 1512 \alpha H^4 = \kappa^2 \rho \\
  \label{cons2} \rhodot + 9 H ( \rho + p ) = 0 \ ,
\eea

\noindent where $H\equiv\adot/a$. The conservation law of the
energy-momentum tensor \re{cons2} is the usual one. But the
Hubble law has qualitatively new features if $\alpha<0$
(which we assume from now on). (For string theory, the second order
coefficient $\alpha$ is, to leading order, zero for superstrings,
and positive for heterotic string theory. However, this is
not the case for all higher order Lovelock terms
\cite{Tseytlin:1995, Maeda:2004}.)

The Hubble law \re{Hubble} is plotted in Figure 1,
along with the usual FRW Hubble law for comparison. The Gauss-Bonnet
Hubble law has two branches, with different vacua and different
dynamics. On branch I the vacuum is Minkowski space, whereas
on branch II the vacuum is de Sitter space with Hubble
parameter $H=1/\sqrt{42|\alpha|}$. The vacua have been analysed
in \cite{Boulware:1985, Myers:1988, Whitt:1988}. In the de Sitter
vacuum, the gravitational excitations are ghosts, implying that
it is not a stable solution.

On branch I, the behaviour is the usual FRW one at low energies
($\kappa^2\rho\ll1/|\alpha|$), with modifications at high energies.
For matter satisfying $\rho+p>0$, the Hubble parameter decreases.
In contrast, on branch II the universe undergoes superinflation
($\Hdot>0$) if the matter obeys $\rho+p>0$: the smaller the
energy density, the faster the expansion of the universe. Likewise,
a positive cosmological constant decreases the expansion rate,
instead of increasing it.

On both branches, the energy density and all other observables are
non-divergent at all times: upon approaching
what would be a curvature singularity in the FRW case, the energy
density levels off. The usual singularity
theorems of general relativity do not apply to Gauss-Bonnet
gravity, so it would be possible for the spacetime to be
non-singular. (If the Gauss-Bonnet tensor is considered as an
effective energy-momentum tensor, it violates the null energy condition.)
However, even though there is no curvature
singularity, the spacetime is geodesically incomplete
and thus singular (\cite{Wald:1984}, page 212).
An easy way to see this is to consider a collapsing universe
on branch I: as the energy density increases to the value
at the peak, $\kappa^2\rho=3/(14|\alpha|)$, the universe
cannot collapse further and simply ceases to exist.

\FIGURE[t!]
{
\includegraphics[width=0.45\textwidth]{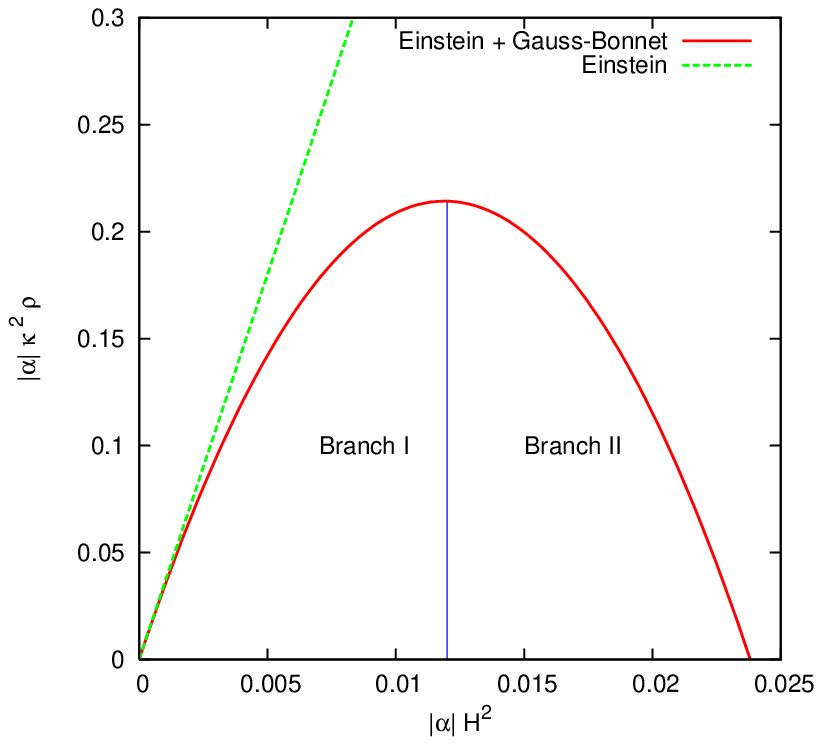}
\caption{The Hubble law with the Gauss-Bonnet term,
and the Einstein Hubble law for comparison.}
}

That the Gauss-Bonnet term leads to an inflationary
solution when $\alpha<0$ can be understood in the following way.
If a tensor does not contain higher than second order
derivatives and is covariantly conserved,
its $00$-component cannot contain higher than
first order derivatives. (Considering the Gauss-Bonnet tensor
as an effective energy-momentum tensor, one can see this
from the conservation law \re{cons2}: if $\rho$ had second
order derivatives, $p$ would be third order.) Given a tensor with
dimension $m^4$ and assuming the spatially flat FRW metric,
the $00$-component is then proportional to
$H^4$, the only available quantity of the correct dimension.
This leads immediately to the Hubble law \re{Hubble}; only
the coefficients 36 and 1512 depend on the detailed structure of the
Gauss-Bonnet tensor (and are specific to 10 dimensions).

The structure is the same for all dimensions $d>4$ where the
Gauss-Bonnet tensor is non-zero. If one includes all
the higher order Lovelock tensors, the Hubble law \re{Hubble}
becomes of order [d/2]-1 with respect to $H^2$. For
$d=10$, there are three Lovelock terms (in addition to the
cosmological constant and the Einstein tensor term), so the Hubble law
is quartic in $H^2$. As long as the Hubble law has at least one non-zero
solution for $\rho=0$, there is at least one inflationary branch.

As an aside, we note that this structure
can be realised even in four dimensions. If the metric is
conformally flat, there exist additional $d/2$
(rounded down) tensors of increasing dimensionality
which are second order in derivatives of the metric
and covariantly conserved \cite{Meissner:2000}. The tensor which
has dimension $m^4$ is usually labeled $H^{(3)}_{\mu\nu}$
(\cite{Birrell:1982}, page 183).
Including this tensor and taking the FRW metric leads to a
Hubble law of the form \re{Hubble}. Some of the properties of the
modified Hubble law discussed above have been earlier
mentioned in this four-dimensional context \cite{Wald:1978}.
The tensor $H^{(3)}_{\mu\nu}$ can even be extended to first order
in perturbation theory around the FRW background \cite{Horowitz}.
Lovelock's theorem guarantees that there is no local extension
of the tensor $H^{(3)}_{\mu\nu}$ to four-dimensional general
spacetimes, but there is a non-local extension
(which is no longer second order in the derivatives),
connected with the trace anomaly \cite{Mazur:2001}.

\paragraph{Graceful exit.}

In order for the inflationary mechanism to be cosmologically relevant,
there has to be a way of ending inflation
--in our case, getting from branch II to branch I--
as well as sorting out only three spatial dimensions to inflate.
In fact, the problems of ending inflation and breaking isotropy
are related. The Hubble law \re{00} in the general case
with $a\neq b$ reads
\bea
  3 ( 1 + 6\lambda + 5\lambda^2 ) H^2 - 36 \lambda ( 2 + 15\lambda + 20\lambda^2 + 5\lambda^3 ) |\alpha| H^4 = \kappa^2 \rho \ ,
\eea

\noindent where $H\equiv\adot/a$, and $\lambda(t)\equiv(\bdot/b)/H$.
If the evolution given
by the Hubble law and \re{yy}, \re{cons} is such that
$\lambda\rightarrow0$, the extra dimensions stabilise,
the Hubble parabola straightens out and branch II disappears.
In other words, inflation ends and the standard Hubble law is recovered.
This happens only if the number of dimensions which
are left free to expand is at most three. For $p$ large
spatial dimensions and $9-p$ extra dimensions, the Hubble law would be
\bea
  && \left[ \frac{1}{2} p (p-1) + p (9-p) \lambda + \frac{1}{2} (9-p) (8-p) \lambda^2 \right] H^2 - \left[ \frac{1}{2} p (p-1) (p-2) (p-3) \right. \el
  && + 2 p (9-p) (p-1) (p-2) \lambda + 3 p (p-1) (9-p) (8-p) \lambda^2 + 2 p (9-p) (8-p) (7-p) \lambda^3 \el
  && \left. + \frac{1}{2} (9-p) (8-p) (7-p) (6-p) \lambda^4 \right] |\alpha| H^4 = \kappa^2 \rho \ .
\eea

\noindent If the extra dimensions are stabilised, $\lambda=0$, we obtain
\bea
  && \frac{1}{2} p (p-1) H^2 - \frac{1}{2} p (p-1) (p-2) (p-3) |\alpha| H^4 = \kappa^2 \rho \ .
\eea

\noindent It is transparent that inflation persists unless
the number of large dimensions is at most three. Note that,
for non-zero $\rho$, stabilisation is not consistent with zero
or one large dimensions. However, there is no obvious obstruction
to having two large dimensions instead of three.
This is a constraint on inflation in the visible
dimensions, assuming that the extra dimensions stabilise.
(If only higher order Lovelock terms were present instead of the
Gauss-Bonnet term, the number of preferred dimensions would be
larger. For the third order Lovelock term, inflation would end for five
or less spatial dimensions, and the fourth order term would yield
seven or less.)

So, if there is a mechanism which allows only a three-dimensional
subspace to become large and slows down expansion of the other
dimensions, the universe will become effectively four-dimensional
and the contribution of the Gauss-Bonnet tensor in the visible
directions will go to zero, ending inflation. Finding such
a mechanism was the original aim of SGC
\cite{Brandenberger:1989}. It is not clear whether three dimensions
are dynamically preferred or not
\cite{Sakellariadou:1995, Easther:2002b, Bassett:2003, Easther:2003, Easther:2004, Danos:2004}.
But even if three dimensions are selected in a slowly expanding
space, with the extra dimensions stabilised by a gas of strings,
in an inflating space the string gas will be diluted and the extra
dimensions will be free to expand \cite{Ferrer:2005, Patil:2004}.
Such a destabilising effect is present even in a matter-dominated
universe, though in that case the string gas can counter the effect
and rein in the extra dimensions \cite{Ferrer:2005}.

We studied whether this stabilisation mechanism with a
gas of strings or higher-dimensional branes could
end Lovelock inflation. We used the energy-momentum tensor
for the string gas given in \cite{Ferrer:2005}, and its
generalisation for higher-dimensional branes.
While strings indeed slow down the expansion initially,
their effect is soon diluted to negligible levels by inflation.
Since the energy density of higher-dimensional branes is diluted
more slowly, they could potentially have a stronger impact.
However, the behaviour is essentially the same:
the brane gas does slow down the expansion of the extra
dimensions, but the effect is too weak, and space
isotropizes, with all dimensions growing large.

So, while we have connected the end of inflation with
(at most) three spatial dimensions becoming large, we have
not managed to explain why the other dimensions would be
stabilised. In the next section, we will discuss some ideas
towards ending inflation and getting from the inflationary
branch to the FRW branch.

\section{Discussion}

\paragraph{Ending and starting inflation.}

The line of reasoning leading to Lovelock gravity (writing down
all terms consistent with second order equations of motion)
is motivated for a classical theory. However, it may be inadequate
when quantum effects are included, because anomalies can break
classical symmetries, leading to a modification of the low energy
action. In the case of quantum fields coupled to classical gravity,
the trace anomaly leads to terms higher than second order in the
derivatives, and one can argue that they should be included in the
effective action of gravity \cite{Birrell:1982, Mazur:2001}.
The terms related to the trace anomaly were used in the first
inflationary model \cite{Starobinsky:1980}. It would be interesting
to investigate their impact on Lovelock inflation.
In particular, the trace anomaly terms could destabilise
the de Sitter solution and lead to a graceful exit,
like in \cite{Starobinsky:1980}.
Like the Lovelock terms, the trace anomaly is
sensitive to the number of dimensions, though it is not clear that it
would prefer three large dimensions over some other number.

From the string theory point of view, the most conspicuous missing
ingredient is the dilaton. We have simply assumed that the dilaton
is stabilised in a way which does not impose constraints on the
Einstein equation. In general, if we include the dilaton in the
action, we have in addition to the Einstein equation the dilaton
equation of motion. Taking the dilaton to be constant then leads
to a constraint equation for the metric.
In the present context with the Lovelock terms, the constraint
removes the de Sitter solution, leaving only the Minkowski
vacuum (somewhat like in the inflationary scenario of \cite{Ellis:1998}).
This might work well, since it means that any period of
inflation would be transient, and the dilaton could serve to
end inflation and take the universe to the FRW branch.
However, while this would tie the end of inflation with dilaton
stabilisation, there is no apparent connection to having three large
spatial dimensions.

Apart from the trace anomaly or dilaton, the fact that the
gravitational excitations around the de Sitter solution are ghosts
implies that it is unstable \cite{Boulware:1985, Myers:1988}.
Such an instability could also provide
a satisfactory transition to the FRW branch.

One advantage of Lovelock inflation is that it is not inconsistent
with a mechanism that would solve the cosmological constant problem
by  cancelling the gravitational effects of vacuum energy,
unlike usual scalar field models of inflation \cite{Brandenberger:1997}.
(For an inflationary mechanism which is instead based on this kind of a
cancellation mechanism, see \cite{Woodard}.)

Another problem of conventional scalar field models 
is getting inflation started. Unless the null energy
condition is violated, starting inflation requires homogeneity
over at least a Hubble-sized patch \cite{Trodden}.
As we have noted, the Lovelock tensors (considered
as an effective source) violate the null energy condition, so
there is no obstruction, in principle, to inflation starting in an
inhomogeneous patch and creating homogeneity, rather than simply
amplifying it. Studies of inhomogeneous spacetimes would be needed
to establish how this works quantitatively; the issue is not fully
worked out even in the usual inflationary case \cite{Ellis}.

\paragraph{Conclusion.}

In the usual formulation of string theory, six spatial
dimensions are compactified by hand, whereas three are taken
to be large. Since the most successful scenario of the early
universe, inflation, produces exponentially large dimensions
starting from small ones, it seems elegant to combine inflation
and the question of why some dimensions are much larger than
others. In this framework, all dimensions would start at some
small natural scale, and inflation would explain why
three of them inflate to become macroscopically large.

We have discussed how a natural generalisation of Einstein
gravity in higher dimensions, Lovelock gravity, can give
inflationary solutions. The inflation will end if one stabilises
the extra dimensions, since the non-Einstein Lovelock terms
do not contribute in 3+1 dimensions or less.
This ties the graceful exit problem of inflation to the number
of spatial dimensions: Lovelock inflation will only end if the
number of large spatial dimensions becomes three or less.

String gas cosmology supplies a mechanism for selecting
only three dimensions to expand. However, while this
mechanism works during both the radiation- and matter-dominated
eras, it fails for inflation. Taking into account the trace anomaly
or the dilaton could lead to a viable graceful exit, but it is not clear
whether the number of large spatial dimensions would emerge correctly.
Further work is needed on stabilising extra dimensions:
what we have shown is that the solution of the stabilisation issue
may be directly relevant for inflation.

\ack

SR thanks Antonio Padilla for discussions, Kari Enqvist and
Esko Keski-Vakkuri for discussions in the early stages
of this work and the Helsinki Institute of Physics for
hospitality. FF is supported in part by grants from the DOE and NSF
at Case Western Reserve University. \\

\appendix

\setcounter{section}{1}

\end{document}